\documentclass[11pt]{article}
\usepackage[margin=1in,footskip=0.75in]{geometry}
\usepackage[utf8]{inputenc}
\usepackage{graphicx}
\usepackage{fbb}
\usepackage{rotating}
\usepackage{soul}
\usepackage{color}
\usepackage{eulervm}
\usepackage{authblk}

\usepackage[colorlinks,citecolor=blue,linkcolor=red,urlcolor=violet]{hyperref}

\newcommand{\Ecol}{\textit{Escherichia coli}}

\newcommand{\etal}{\textit{et~al.}}

\title{Designing biological circuits: from principles to applications}
\author[1,2,3]{Debomita Chakraborty}
\author[2,3,4]{Raghunathan Rengaswamy}
\author[1,2,3]{Karthik Raman\thanks{Corresponding author. E-mail: kraman@iitm.ac.in}}
\affil[1]{Bhupat and Jyoti Mehta School of Biosciences, Department of Biotechnology, Indian Institute of Technology (IIT) Madras, Chennai - 600 036, India}
\affil[2]{Centre for Integrative Biology and Systems mEdicine (IBSE), IIT Madras, Chennai --- 600 036, India}
\affil[3]{Robert Bosch Centre for Data Science and Artificial Intelligence (RBCDSAI), IIT Madras, Chennai --- 600 036, India}
\affil[4]{Department of Chemical Engineering, IIT Madras, Chennai --- 600 036, India}

\date{}

\begin{document}

\maketitle

\begin{abstract}
Genetic circuit design is a well-studied problem in synthetic biology. Ever since the first genetic circuits---the repressilator and the toggle switch---were designed and implemented, many advances have been made in this area of research. The current review systematically organizes a number of key works in this domain by employing the versatile framework of generalized morphological analysis. Literature in the area has been mapped based on (a)~the design methodologies used, ranging from brute-force searches to control-theoretic approaches, (b)~the modelling techniques employed, (c)~various circuit functionalities implemented, (d)~key design characteristics, and (e)~the strategies used for the robust design of genetic circuits. We conclude our review with an outlook on multiple exciting areas for future research, based on the systematic assessment of key research gaps that have been readily unravelled by our analysis framework.
\end{abstract}

\section{Introduction}
The design of biological circuits capable of achieving specific functionalities is a cornerstone of synthetic biology. These circuits comprise various components, such as genes, promoters, transcription factors (TFs), and proteins, for implementing functionalities with biological significance. The idea of genes acting as \emph{biological circuits} finds its origins in the seminal work on the \emph{lac} operon, by Jacob and Monod~\cite{jacob_genetic_1961}. As a result of this `circuit', a cell preferentially takes up glucose when there is a simultaneous presence of lactose and glucose. The cell achieves this by \emph{switching off} the gene that codes for lactose as long as glucose is available.

In 2000, two classic biological circuits---the repressilator~\cite{elowitz_synthetic_2000}, and a toggle switch~\cite{gardner_construction_2000}---were the first genetic circuits to be designed using mathematical models and implemented \textit{in vivo} in \Ecol{} using various genetic constructs.  

A key aspect of synthetic biology is the application of engineering systems principles for the rational design of biological circuits using a bottom-up or reductionist approach. However, certain inherent properties of biological systems pose significant challenges to such a design approach. For instance, biological components, though modular in their own right, are nowhere as modular as engineered systems---such as those comprising electrical components. Therefore, it is a common occurrence that the interconnection of even simple, well-characterized components to build more complex circuits, leads to deviations from expected behaviours. Such deviations may be attributed to the property of weak emergence in biological systems~\cite{Bedau2013}.

\begin{figure}
      \centerline{\includegraphics[width=\linewidth]{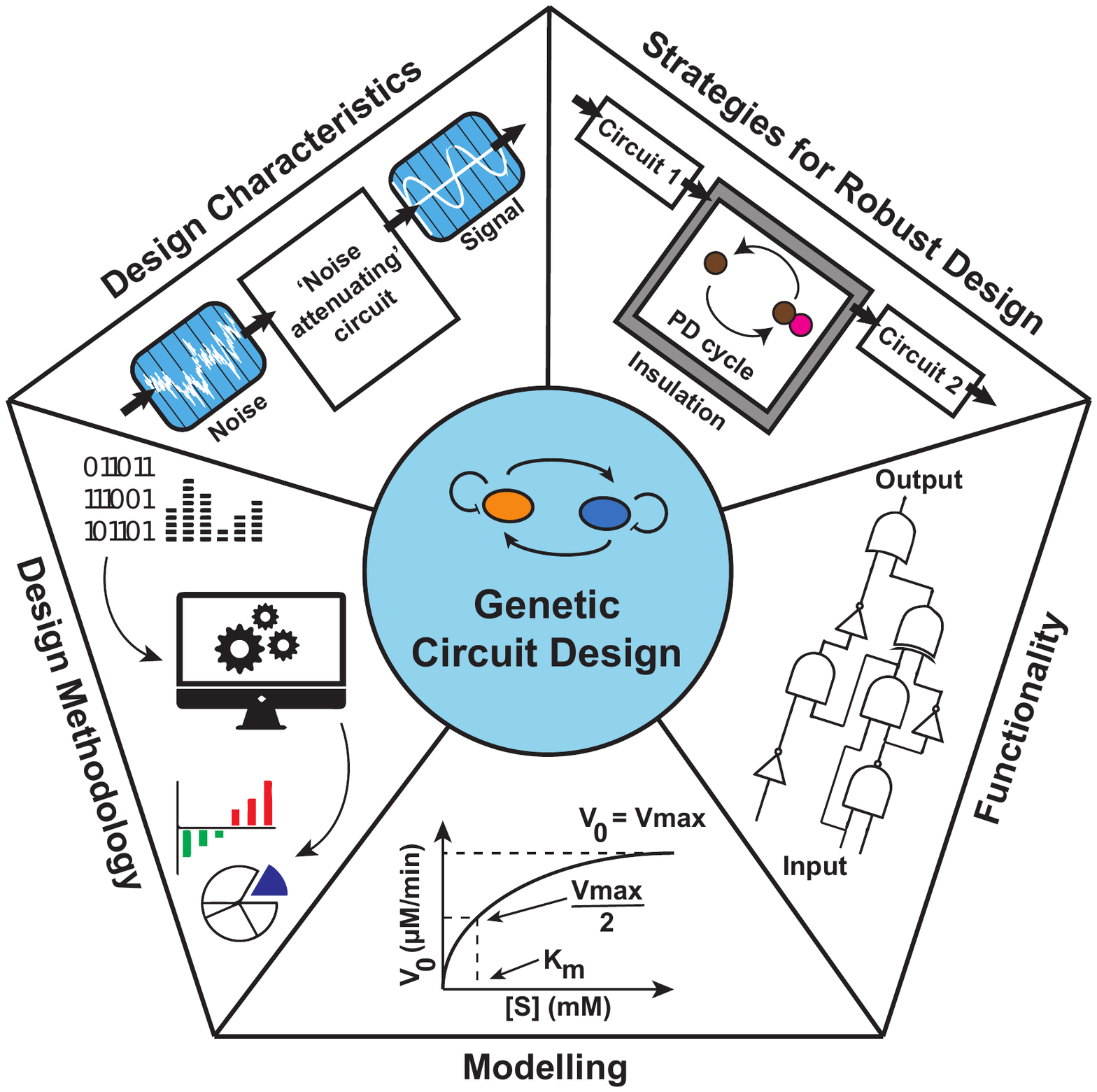}}
      \caption{Key \emph{parameters} underlying the genetic circuit design problem. (i) Design methodology ($\S$\ref{sec:design_methodologies}), (ii) functionality ($\S$\ref{sec:functionality}), (iii) modelling ($\S$\ref{sec:modelling}), (iv) design characteristics ($\S$\ref{sec:design_characteristics}), and (v) strategies for robust design ($\S$\ref{sec:strategies_robust_design}) were identified as the key \emph{parameters} into which the reviewed literature can be classified. Using these \emph{parameters}, the generalised morphological analysis table (Table~\ref{tab:gma}) was constructed.}
      \label{fig:gmadiag}
    \end{figure}

Several researchers are working to unravel design principles that underlie reliable biological circuits. In this  review, we systematically classify existing research in this domain into various streams and identify potential research gaps that remain to be addressed. For this purpose, we apply the framework of generalised morphological analysis~\cite{ritchey_general_2018} to lay out the existing literature on biological circuit design. Generalised morphological analysis is a method for analysing a complex, qualitative problem by identifying the problem's critical parameters. Figure~\ref{fig:gmadiag} shows five key parameters underlying the genetic circuit design problem. The framework lists a set of ``options'' for each ``parameter'' that characterises a problem (Table~\ref{tab:gma}). Subsequently, constructing a cross-consistency matrix (CCM) helps unravel associations across different parameters and options. Using the generalised morphological analysis framework and the CCM, it is possible to get insights into the existing knowledge about a problem and identify potentially unexplored areas. Thus, it is useful for systematically finding future research directions in a given domain of interest.

The rest of this manuscript is organised as follows: Table~\ref{tab:gma} is central to the manuscript and shows a generalised morphological analysis of the literature reviewed, and Table~\ref{tab:ccm} shows the corresponding CCM. Section~\ref{sec:design_methodologies} lists various genetic circuit design methodologies; Section~\ref{sec:modelling} lists some ways to mathematically model genetic circuits; Section~\ref{sec:functionality} describes different circuit functionalities built by researchers. Section~\ref{sec:design_characteristics} lists the characteristics that the designs address. Section~\ref{sec:strategies_robust_design} discusses the strategies found for robust design. The last Section~\ref{sec:outlook} presents a synthesis of our observations on the key research gaps identified from literature and future perspectives.

\begin{table}[htp]
    \centerline{\includegraphics[width=\linewidth]{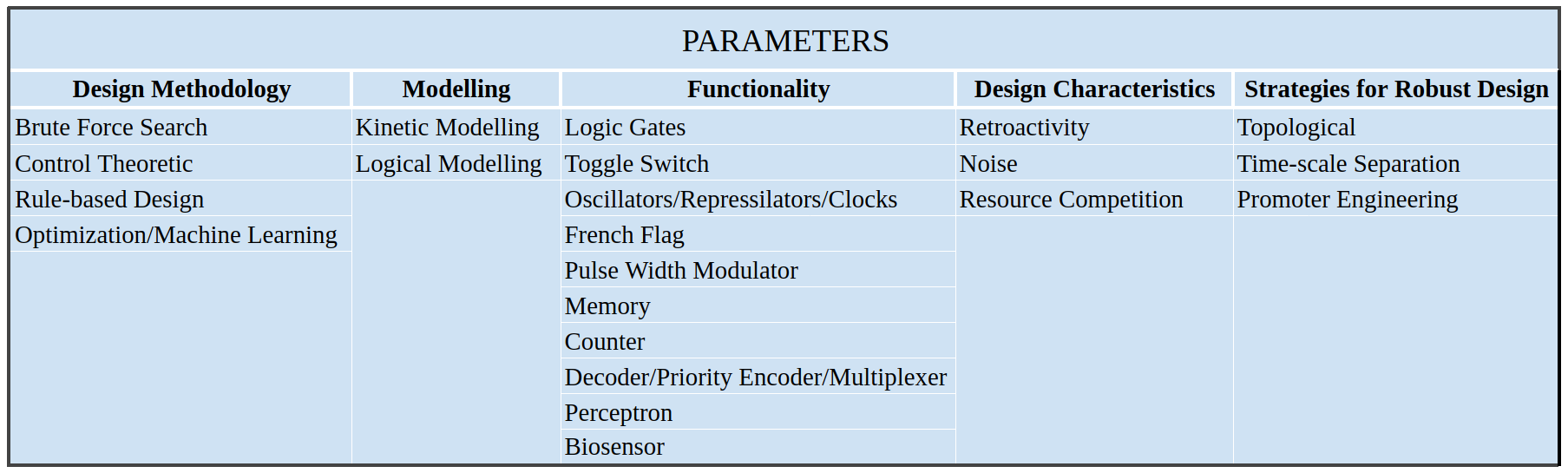}}
    \caption{\textbf{Generalized Morphological Analysis of literature in the domain of genetic circuit design.} Each column in the table represents the \emph{parameters} underlying the problem of genetic circuit design. The topics listed under each \emph{parameter} are called the corresponding \emph{options}. These \emph{parameters} and \emph{options} are integral to the generalised morphological analysis, and are used to construct the matrix in Table~\ref{tab:ccm}.}
    \label{tab:gma}
\end{table}

\begin{sidewaystable}[htp]
    \includegraphics[width=\linewidth]{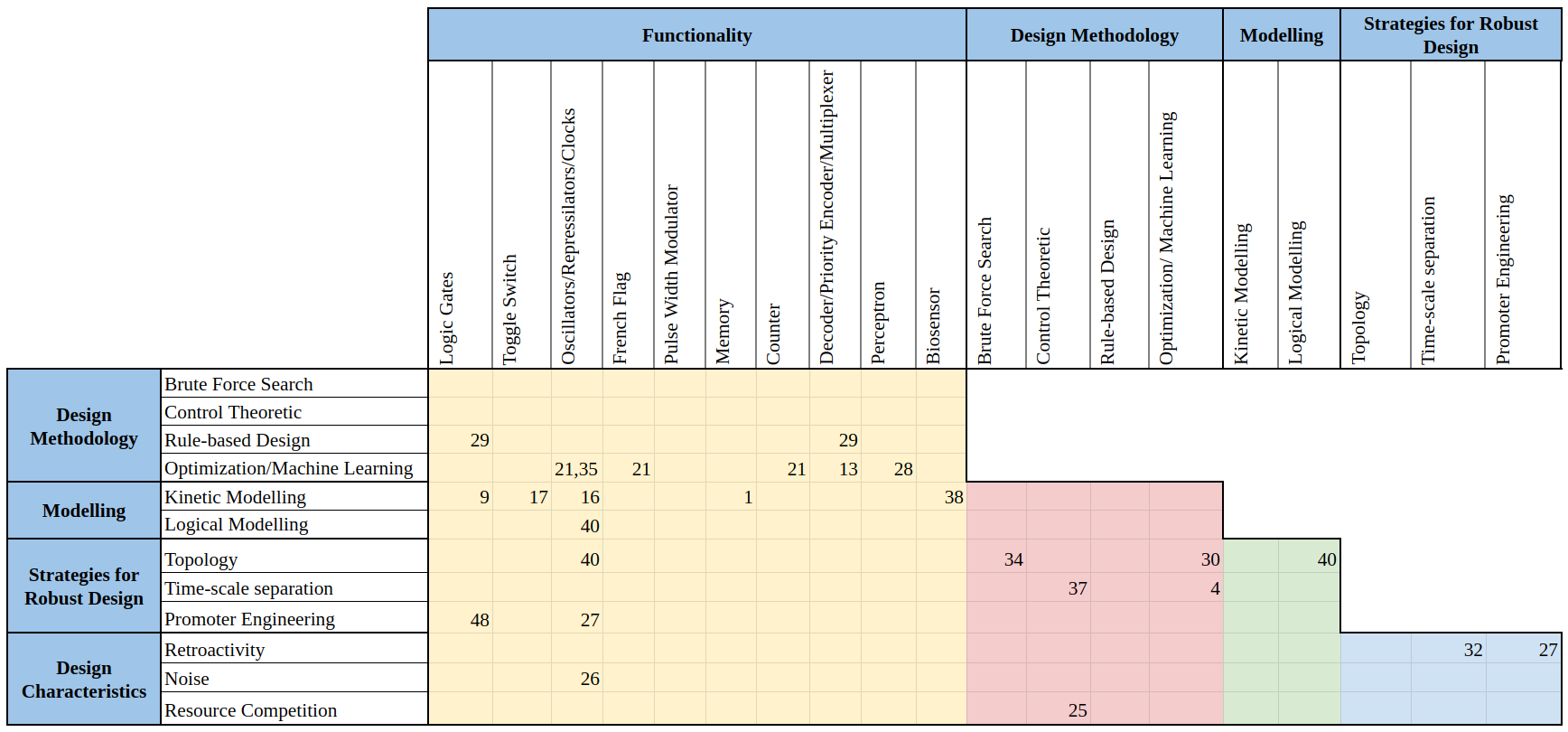}
    \caption{\textbf{Cross-consistency matrix (CCM) for the literature reviewed.} The rows and columns of the CCM are derived from the \emph{parameters} and \emph{options} given in Table~\ref{tab:gma}. Each cell in the matrix represents the intersection of the topics of the corresponding row and column. Numbers indicate references dealing with the work done on the topic represented by the corresponding cell. The empty cells represent topic combinations that are hitherto unexplored, or not covered under this review; they could also represent combinations that are infeasible/impossible. }
    \label{tab:ccm}
\end{sidewaystable}

\section{Design Methodologies}
\label{sec:design_methodologies}

Multiple design methodologies have been employed for the construction of genetic circuits. These approaches range from brute force searches of circuit design space, to optimisation-based approaches, to ultimately control-theoretic and rule-based approaches, which are rooted in a deeper understanding of circuit properties, and how they relate to topology. 
The following subsections discuss each of these methodologies.

\subsection{Brute Force Search}
A comprehensive search constitutes the simulation of all possible topologies with a predefined number of nodes to identify the topologies that achieve a function or exhibit a phenotype of interest. A map of the topology--function relationship has been obtained for adaptive motifs in three-node enzyme networks using the brute force search~\cite{ma_defining_2009}. In this work, the authors simulated 16038 circuits and calculated their sensitivity and precision in achieving adaptation over 10000 parameter sets. Out of the simulated circuits, 395 showed adaptation, with only two motifs recurring in these adaptive networks. This work demonstrated the power of computational methods for getting insights into the design principles of enzyme networks.

Tang and co-workers~\cite{shi_adaptation_2017} subsequently did a similar analysis for adaptation in three-node transcription regulation networks (TRNs).  Like enzyme networks, the TRNs also have only two recurring motifs in all three-node networks that exhibit adaptation. These motifs are called the Negative Feedback Loop with an Exponential Buffer node (NFBLEB) and the Incoherent Feed-Forward Loop with an Inversely Proportional node (IFFLIP).

Despite the insights drawn from a brute force search method, it is not a computationally scalable design methodology. A systems-theoretic approach~\cite{bhattacharya_systems-theoretic_2018} has led to similar conclusions as~\cite{ma_defining_2009}. Another technique called TopoFilter~\cite{lormeau_multi-objective_2017} uses Approximate Bayesian Computation (ABC) for topological filtering and samples the parameter space for the viability of three-node enzyme networks for multiple objectives: robustness, feasibility, and performance. This work, too, led to the same conclusions as~\cite{ma_defining_2009} but without requiring an exhaustive search. TopoFilter identified the viable motifs for adaptation and explored the entire parameter space to identify feasible regions for each adaptive circuit. Various trade-offs in the multiobjective design problem were identified by correlating the feasible parameter spaces for pairwise combinations of the objectives. It is found that robustness and feasibility are orthogonally related, while a more robust circuit shows a lower performance. Amongst robust circuits, the ones having simpler topologies exhibited a higher degree of robustness than those with complex topologies~\cite{lormeau_multi-objective_2017}. The results reported are for enzyme networks, and the same for TRNs are yet to be studied. Recently TopoFilter has been made available as a MATLAB package for mechanistic model selection~\cite{rybinski_topofilter_2020}.

\subsection{Control-Theoretic Approaches}
The classical design principles of engineering systems may be applied to design synthetic biological circuits. Biological circuits are inherently non-linear, stochastic, and composed of parts such as genes, transcription factors, and proteins that are not modular. This lack of modularity leads to inconsistent behaviour upon interconnection of parts. Hence, employing well-known control-theoretic measures for handling non-linearity, noise, and loading effects in the design of synthetic genetic circuits is a promising strategy. Negative feedback control is used extensively in the robust design of engineering systems. In synthetic biological circuits, there are two ways to implement negative feedback control: (i)~in-cell feedback control---the controller is implemented within the cell, and (ii)~\emph{in silico} feedback control---an extracellular controller controls a group of cells~\cite{del_vecchio_control_2016}.

In another study~\cite{rivera-ortiz_integral_2014}, the efficacy of integral feedback control in achieving a robust design of synthetic biological circuits is highlighted. An integral feedback controller attenuates the error between the output and the reference input by taking the error's time integral. The error can be reduced to zero at a steady state if the reference input and the error are constants. Oscillatory genetic circuits with regular and chaotic oscillations are often found in biochemical systems, such as circadian clocks. The effect of integral control in such oscillatory systems is studied while presenting a generalised way to define homeostasis~\cite{thorsen_effect_2019}. For instance, oscillatory or chaotic systems that maintain an average value of regulated molecules rather than a constant steady-state value may be considered homeostatic. Baetica \etal~\cite{baetica_control_2019} call for an updated definition of homeostasis as the concept of a reference input may not hold good in biochemical systems. It is argued that the block diagram representation used in control theory essentially assumes the modularity of components. This assumption does not represent the actual scenario for many biochemical systems. Thus, there is a need to find novel ways to adapt control-theoretic methods for the design of synthetic biological systems.

Tangirala and co-workers~\cite{Bhattacharya2021generic} have proposed a generic systems theory-driven approach to design protein networks capable of perfect adaptation. They use systems theory to obtain mathematical constraints from the necessary qualitative conditions for adaptation; these constraints further provide design requirements for the underlying networks, unravelling key design principles.

\subsection{Rule-based Design}
The design of genetic circuits requires extensive knowledge of the underlying biology, including understanding specific components like promoters, which are different for different organisms. Other components, such as the genetic code for a particular protein, are the same across different species. Moreover, a design intended for a particular context may not be reusable in another context. This lack of portability of a design poses a significant bottleneck in the fast and scalable design of synthetic biological circuits even though the technologies to implement the design are already available. Many researchers believe that genetic parts should be characterised and recorded in a standardised library to facilitate a more viable design cycle~\cite{Drubin2007}. At the same time, the domain experts can translate their knowledge into design rules in context-free grammar (CFGs)~\cite{cai_syntactic_2007}. Such an approach is called rule-based design and is used in the bio-design automation tool GenoCAD~\cite{wilson_step-by-step_2011}.

Another genetic design automation software called Cello is introduced by Nielsen \etal~\cite{nielsen_genetic_2016}. Cello uses a hardware description language (Verilog) to design a biological circuit like electronic circuit design. Such designs are independent of the actual biological parts to be used in the circuit implementation. The specifications of the biological parts or required topological motifs for circuit implementation, the circuits' operating conditions are all defined in a User Constraints File (UCF). During logic synthesis, the circuit design is subjected to the rules defined in the UCF. A Monte Carlo simulated annealing algorithm then handles the optimal way to interconnect the circuit's logic gates. This method allows the implementation of the same circuit functionality using different biological parts from a library of standard parts (such as BioBricks or iGEM). To handle the problem of retroactivity, i.e. the change/loss of modules' functionality upon interconnection, the modules are separated by insulation.

\subsection{Optimisation/Machine Learning (ML)}
Hiscock proposed an approach that combines ODE-based modelling and optimization~\cite{hiscock_adapting_2019}. A fully connected network of a particular number of nodes is modelled using a system of coupled ODEs. The functionality of interest (e.g., oscillation, pulse detection) is identified for which the network parameters are to be optimized. The ODE model consists of a matrix parameter that determines which of the nodes in the network interact. One of the nodes is associated with the input parameter which is translated by the network to another node which acts as the output node. Based on the functionality of interest, the desired input-output node relationship is determined. An ODE solver generates the desired input-output node concentration data. An optimization problem is subsequently formulated with the mean-squared error between the desired output and the actual output as the cost function. The desired parameter set is then searched in the high dimensional parameter space by minimization of the cost function. The search method used to find the parameters is the advanced gradient descent algorithm, Adam. This algorithm is available as a Python module, GeneNet. It has a better speed than evolutionary algorithms and comprehensive search through parameter space for comparable networks. It is also shown to be scalable for circuits with up to nine nodes. The complexity of the network can be controlled using an additional regularisation term that takes the L1 norm of the parameters in the cost function. However, it finds only one suitable circuit that implements a required functionality and does not provide alternative circuits that might be relevant. 

Smith \etal~\cite{smith_designing_2017} have employed an evolutionary algorithm to find the optimal circuit that exhibits a particular functionality of interest. Another optimisation framework, OptCircuit~\cite{dasika_optcircuit_2008}, is used to design genetic circuits and to tune parameters for better performance although the design of complex circuits using this framework is computationally expensive.

\section{Modelling}
\label{sec:modelling}

\subsection{Kinetic Modelling}
The operation of a genetic circuit is driven by transcription of synthetically designed genetic parts. Dynamical modelling of transcription typically assumes the cell's biochemical environment to be well-mixed or homogeneous in space. Therefore, a set of ODEs generally models the chemical kinetics underlying transcription. The most common kinetic models of transcription are one of the following: 

\subsubsection{Mass-action Kinetics} 
The law of mass-action states that the reaction rate is proportional to the probabilities of collisions between the molecules of the reacting species~\cite{guldberg1879}. It is widely used to model transcription dynamics~\cite{rivera-ortiz_integral_2014}.

\subsubsection{Hill Kinetics}

The Hill function is also widely used to model transcription dynamics~\cite{shi_adaptation_2017,zong_insulated_2017}. The Hill function is a sigmoidal function that becomes more switch-like with increasing the cooperativity of TFs in binding to a promoter to activate/repress a gene. In Hill function-based models, it is assumed that the TFs in the system simultaneously bind to the promoter. The model is based on time-scale separation, similar to Michelis-Menten equation-based models for enzyme networks. The binding of TFs is considered a fast process that quickly reaches equilibrium. In contrast, the expression of the target gene is a slow process~\cite{gunawardena_time-scale_2014}.

\subsubsection{Generalized Mass Action}

The reactions that constitute transcription are non-linear. Several functional forms have been used for ODEs modelling the dynamics of the system. These functional forms are often based on approximations using Taylor's theorem. The simplest approximation is a power-law form that lies at the core of generalised mass action and $S$-system representations~\cite{alves_mathematical_2008}. Power-law terms represent the individual processes in the system. The entire circuit or network is modelled by aggregating these terms for each node.

\subsection{Logical Modelling}
Bernot \etal~\cite{bernot_formal_2014} proposed a logical modelling approach for gene regulatory networks based on R. Thomas' idea of discrete modelling. In this approach, the continuous space of protein concentrations is translated into a discrete phase space by marking intervals based on thresholds at which interaction between a pair of proteins happens. The exact values of the thresholds may be unknown. The components in a genetic circuit and the type of interactions amongst them, such as activation, repression, or no effect, need to be only qualitatively known. Given this information, a circuit operation can be translated into simple logical structures like AND, OR, and NOT. There may be different conditions under which other logical functions get executed. For example, suppose a gene should be expressed in the absence of a repressing protein provided specific activating TFs are present. Alternatively, suppose the gene should be expressed unconditionally (i.e., irrespective of activating proteins) in the absence of the repressing protein. Any such set of conditions can be represented using multiplexers. An entire gene network can be represented using multiple levels of multiplexers. This is a static representation of the network. Since the concentrations can only change continuously, not all transitions within the discrete phase space are possible. A notion of neighbourhood arises wherein only the possible transition from one region in concentration space to another region is a valid state change of the network. Ultimately, a dynamical representation of the network is obtained by constructing a state-space graph.

Another design method~\cite{valderrama-gomez_phenotype-centric_2018} combines notions from logical and mechanistic modelling. A circuit is modelled mechanistically  with rate laws representing its underlying biochemical processes. The biochemical reactions are modelled using generalized mass action equations consisting of only algebraic terms of power-law functions. Some parameters in these equations represent fixed features of the system, such as the sign of interaction (\textit{i.e.}, activation or repression), kinetic orders, and the information regarding which species interact. On the other hand, some parameters like the rate constants or environmental inputs are variable.

Each of the defining equations contains ${P}_{i}$ positive, and ${Q}_{i}$ negative terms. The equation for each constituent species of the system would have a dominant positive term (input process) and a dominant-negative term (output process).  A subsystem called S-system is constructed using only the dominant processes for each of the constituents. The equations representing the S-system are linear in logarithmic space since they contain only power-law terms. The Design Space Toolbox~3 (DST3)~\cite{valderrama-gomez_mechanistic_2020,valderrama-gomez_phenotype-centric_2018} constructs all possible S-systems from the defining equations and calculates their eigenvalues. Each eigenvalue is associated with a phenotype. The S-system corresponding to this eigenvalue provides the required network. The system exhibits the phenotype over a polytope in parameter space bounded by linear hyperplanes (in log space). The volume, shape, and boundaries of this polytope define the global robustness of the phenotype. Some possible phenotypes are bistability, hysteresis, oscillations. This method is suited for designing genetic networks since gene-gene interactions are mostly well-defined qualitatively but have many unknown parameters.

\section{Functionality}
\label{sec:functionality}

The design of a genetic circuit contains genes specified as the input and output nodes. The functionality of the genetic circuit is defined as the desired map of the concentration profile of the protein expressed by the output gene over time for a particular change pattern in the concentration of the protein expressed by the input gene. The repressilator~\cite{elowitz_synthetic_2000} and bistable switch~\cite{gardner_construction_2000} were the earliest synthetic genetic circuits designed. A repressilator contains a ring-like network of repressor genes. Due to differences in the rates of repression of the genes, one or more of the genes express the corresponding protein in an oscillatory fashion. Thomas~\cite{thomas_relation_1981} reported the design of repressilators with an odd number of nodes since only circuits with an odd number of repressors could produce sustained oscillations. There have been various experimental implementations of oscillator circuits with linked negative and positive feedback~\cite{stricker_fast_2008}. The timekeeping mechanism in eukaryotes uses circadian clocks, which are basically limit-cycle oscillators, with 24-hour periods~\cite{monti_robustness_2018}.

Genetic circuits with functionalities similar to digital electronic circuits such as logic gates, counters, and decoders, have also been designed. The implementation of NOR gates in the multicellular regime has been demonstrated~\cite{tamsir_robust_2011}. A French flag circuit, a biological counter that can count the number of pulses of different amplitudes or durations, has been designed using GeneNet~\cite{hiscock_adapting_2019}. A genetic field-programmable ROM (FPROM) circuit is built using  Boolean Logic and Arithmetic through DNA Excision (BLADE)~\cite{weinberg_large-scale_2017}. BLADE is a general framework for designing genetic circuits in mammalian cells with recombinases instead of transcription factors. The FPROM circuit allows the user to program the circuit into 16 possible logic gates, which can be single or two-input gates, with the help of some select inputs. Also, more than 100 circuit functionalities, including multiple-input logic gates, have been implemented using BLADE. Ajo-Franklin \etal~\cite{ajo-franklin_rational_2007} have designed and implemented a memory circuit in yeast. In~\cite{pinheiro_synthetic_2016}, the authors reported the \emph{in silico} design of a synthetic gene network that can function as a perceptron capable of associative learning. Tabor \etal introduced an edge-detecting sensor in a community of \Ecol{} cells engineered using an assembly of genetic circuits that sense the edges between light and darkness~\cite{tabor_synthetic_2009}.

A genetic decoder circuit, a toggle switch, logic gates, and a concentration band detector circuit were designed using the optimisation framework, OptCircuit~\cite{dasika_optcircuit_2008}. The genetic circuit design automation tool, Cello, was used to design a priority detector, a consensus circuit with three inputs, and a host of other circuits for various combinatorial logic operations~\cite{nielsen_genetic_2016}. Bagh and co-workers~\cite{bonnerjee_design_2019} have reported the design and fabrication of a three-input AND gate. 
A number of other functionalities such as pulse width modulator and biosensors, as listed in Table~\ref{tab:ccm}, have also been studied, in a variety of systems~\cite{Chuang2014}.

\section{Design Characteristics} 
\label{sec:design_characteristics}

Integrated electronic circuit design methods are often motivated by specific objectives such as the minimisation of power consumption or maximisation of circuit speed based on the applications. Such circuit design methodologies are said to be characterised by the objective it attempts to fulfill, for example, low power design. Similarly, the design methodologies in synthetic biology may be characterised by certain objectives that we term as the \emph{design characteristic}. The choice of the design characteristic is motivated by the demands of the application of the circuit. In this regard, the current focus of genetic circuit designers is to enhance the reliability of genetic circuit designs. However, synthetic biological systems are sensitive to changes in various factors such as temperature, osmolarity, and local redox potential. Their functionalities deviate from the design depending on the context. Such systems are often constructed by assembling subsystems, and the interactions of these systems are also context-dependent. The context-dependent mechanisms that pose a challenge to the modular design of biological circuits are of three types: (i)~compositional context, (ii)~host context, and (iii)~environmental context~\cite{cardinale_contextualizing_2012}. Some of the effects that arise due to context-dependence are retroactivity, noise and resource competition. We define design characteristic as the context dependent effect that is addressed while designing a genetic circuit. 

\subsection{Retroactivity}

The interconnection of a well-characterized synthetic circuit to other components or even a single downstream node can lead to loss of functionality. This effect is known as retroactivity and is an important context-dependent effect that must be addressed for the reliable design of genetic circuits~\cite{Wang2019}. 
A combination of \emph{in silico} and \emph{in vivo} methods have been used to study retroactivity in a dual-feedback oscillator circuit~\cite{moriya_comparison_2019}. 

\subsection{Noise}

Biological systems are inherently noisy, and the stochasticity of the underlying biochemical reactions can make a circuit deviate from its intended design. Natural biological systems are highly robust and tend to maintain their functionality despite operating in the presence of noise.

In~\cite{monti_robustness_2018}, the authors hypothesise that timekeeping systems in cells have evolved to maximise the robustness to input noise. Stoof \etal{} have presented a model of the spatial dynamics of TFs is assuming that TFs perform a one-dimensional local search for their target promoters along the chromosome~\cite{stoof_model_2019}. The TFs undergo diffusion in three dimensions within the cytoplasm. The distance between the gene coding for the TF and the target gene that it expresses is modulated. The noise in the expression of the target gene is found to increase with this intergenic distance. Hence, it opens up the possibility of modelling some noise components in genetic circuits as deterministic phenomena rather than a stochastic process. Therefore it is suggested that intergenic distance may be used as a design parameter for synthetic genetic circuits. 

\subsection{Resource Competition}

A genetic circuit incorporated in a cell shares the cell's in-built transcription machinery. Thus, there is a scope for competition for resources with the existing genes that may affect the circuit's performance. McBride and Del Vecchio~\cite{mcbride_resource_2018} defined metrics for quantifying the resources that a circuit requires and determined the circuit's sensitivity to the availability of resources.

\section{Strategies for Robust Design}
\label{sec:strategies_robust_design}

Robustness is a crucial property of cells, making them less sensitive to environmental changes and internal changes in the concentration of components and mutations. However, the structural features that impart robustness to cellular networks are not yet well-understood. In the context of genetic circuits, robust design refers to design that does not deviate from the intended functionality when connected with other components or placed \emph{in vivo}, in the presence of noise and cross-talk. It is essential to explore design principles that consider various context-dependent effects and effectively insulate a module from them. Some strategies found in the literature for the insulation of genetic circuits include the following: (i)~incorporating topological network motifs such as negative feedback loops that make a circuit inherently robust, (ii) time-scale separation by manipulating the futile cycles of phosphorylation-dephosphorylation (PD), and (iii) promoter engineering.  

\subsection{Circuit-topology based approaches}

The use of graph theory in the study of biological networks is based on the paradigm that a network's function is related to the network topology~\cite{thomas_relation_1981}. Adaptive circuits invariably have specific topological motifs such as negative feedback and feed-forward loops. This topology-function relationship has been validated by several strategies such as brute force search~\cite{shi_adaptation_2017,ma_defining_2009}, systems theoretic methods~\cite{bhattacharya_systems-theoretic_2018}, Approximate Bayesian Computation algorithm for topological filtering~\cite{lormeau_multi-objective_2017} and Mixed Integer Dynamic Optimization~\cite{otero-muras_distilling_2019}. The question that arises is whether there exist topologies that can make genetic circuits inherently robust to retroactivity, noise, or cross-talk. If identified, such topologies would reduce/eliminate the need for separate insulating devices for interconnecting downstream modules to a circuit.

\subsection{Time-scale separation} Different biochemical processes such as transcription factor binding to promoters, phosphorylation of proteins, and molecule-molecule interactions occur at different speeds. The separation in time scales of these processes may be leveraged to insulate various modules of a circuit.  Phosphorylation-dephosphorylation (PD) cycles, GTPase cycles, or phospho-relays introduce futile cycles that insulate a genetic circuit's modules, making it robust to retroactivity. Subsequently, it has been computationally demonstrated that the insulating PD cycles increase the energy cost in ATP and add to the cell's metabolic load~\cite{barton_energy_2013}. A phospho-relay system~\cite{rivera-ortiz_integral_2014} using integral action and a slower time scale for the input signal and the disturbance (caused by the connection of downstream nodes) compared to the time-scale of the system dynamics results in successful attenuation of loading effects. Such phospho-relays are energy-efficient insulating parts analogous to unity gain buffers in electronic circuits.

\subsection{Promoter Engineering} In prokaryotes, the promoters consist of a promoter core for initiation of transcription and operators for transcription factor binding. These components are not entirely independent of each other. They need to be decoupled to make them modular. Promoter engineering has been performed experimentally where promoter cores have been decoupled from the operators to obtain minimal promoters that do not interact with surrounding operators~\cite{zong_insulated_2017}. The minimal promoters were then characterised and the information stored in a database. This database can be referred to in the bottom-up design of promoters that provide insulation when used in a genetic circuit.

\section{Outlook}
\label{sec:outlook}

The design of synthetic genetic circuits is an exciting area of research that has seen much progress in recent times. This is a sub-domain of synthetic biology that uses a reductionist approach to design gene regulatory networks. In incorporating engineering principles into synthetic biology, the process flow to build synthetic genetic circuits may be divided into design and implementation cycles similar to other engineering problems. The design process involves mostly \emph{in silico} modelling of the circuits and performing mathematical or computational tests to come up with designs that work in theory. The implementation process encompasses the wet-lab procedures using the available technologies to actually implement the designs. This review focuses on the research on the design aspects of genetic circuits rather than the implementation technologies. The insights drawn from the study are discussed below.

In this paper, some key methodologies for genetic circuit design have been covered. It is evident that a brute force search for robust design is not a computationally scalable approach. Only up to three-node networks have been simulated using brute force search to identify enzyme networks and TRNs that show adaptation. In contrast, an Approximate Bayesian Computation based algorithm and a system theoretic method arrived at similar conclusions as brute force search but at a lesser computational cost. These methods have been executed for enzyme networks only and are yet to be applied to TRNs. Since biological regulation networks, in their ability to adapt to varying conditions, resemble neural network algorithms~\cite{hiscock_adapting_2019}, ML may be a suitable choice for such network design, especially for realising emergent functions. However, the computational cost of the algorithm for designing circuits must also be scalable for increasing circuit complexity to make it a viable design methodology. A rule-based design approach appears suitable for rapid generation of reproducible designs as it uses standardised libraries of biological components. Such standard components can facilitate commercialisation of synthetic genetic circuits by using the services of bio-foundries~\cite{Hillson2019}. Since rule-based design uses automation software the designs are often simple and the full potential to optimise a given circuit may not be exploited, as can be done in the case of design using control-theoretic methods. Control-theoretic methods are promising, as designs using integral feedback control have revealed~\cite{rivera-ortiz_integral_2014}. The logical modelling approach can also provide useful insights into the design space and the global robustness of particular phenotypes, such as oscillations and steady states~\cite{thomas_relation_1981}. 

\begin{figure}
      \centerline{\includegraphics[width=17cm]{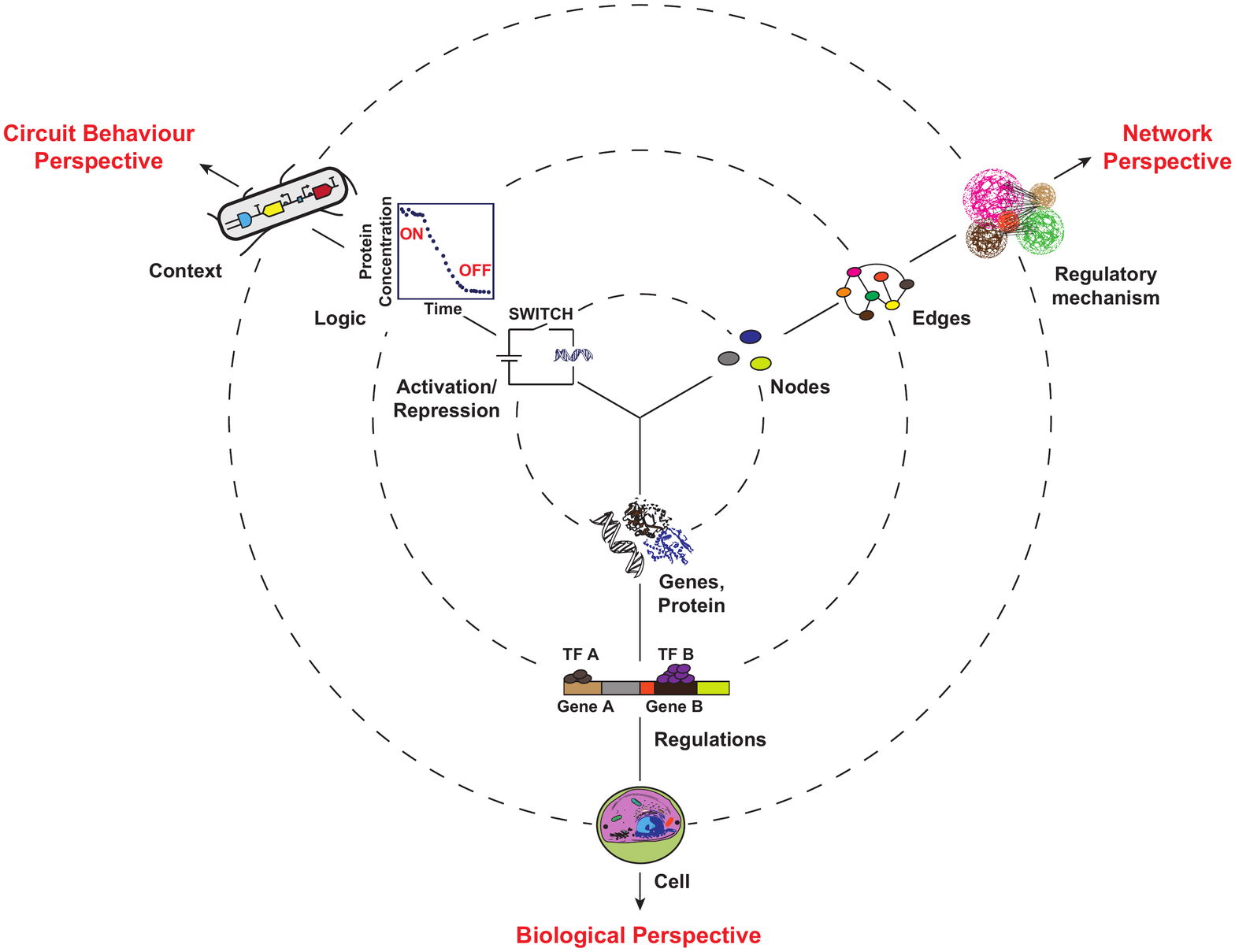}}
      \caption{A Y-chart showing the perspectives in genetic circuit design. The \emph{network perspective} refers to the design of genetic circuit by representing the genes and the regulatory interactions amongst them as nodes and edges, respectively of a directed graphs. The \emph{biological perspective} refers to the actual physical design of genetic circuits using wet-lab processes while the \emph{circuit behaviour perspective} refers to the design approach by relating the regulatory interactions between genes as circuit functions. The actual design of a genetic circuit involves integrating all the three perspectives at different levels of abstractions.}
      \label{fig:ychart}
    \end{figure}

In this review, some of the key works on genetic circuit design have been discussed. Specifically, we systematically map out literature using a generalised morphological analysis, to systematically unravel the parameters into which the studied literature can be classified. Furthermore, the CCM has been constructed to identify potential gaps in the existing work done in this domain and motivate future research directions (Table~\ref{tab:gaps}). A Y-chart (Fig.~\ref{fig:ychart}) based on a similar chart defined for integrated circuit design is constructed to show the problem of genetic circuit design from three different perspectives, viz., the biological perspective, the network perspective that abstracts the circuit into a network representation, and the circuit behaviour perspective showing the functionality taking place at different component levels.
    
\begin{table}
    \centerline{\includegraphics[width=17cm]{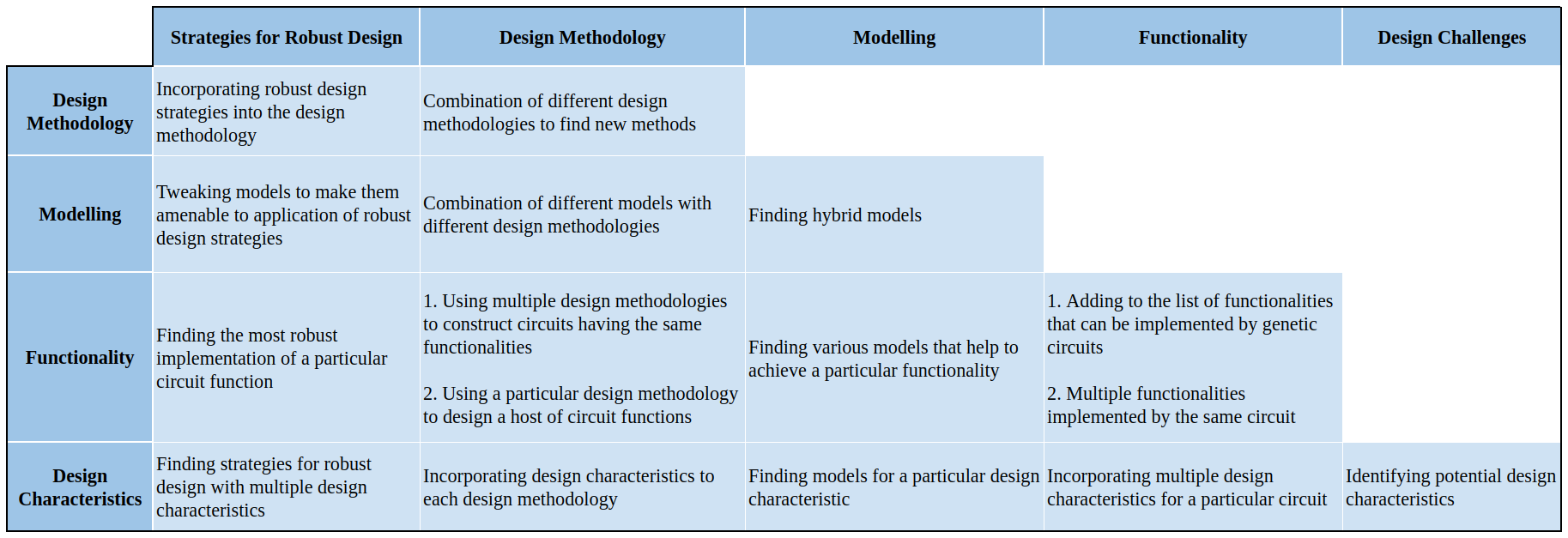}}
    \caption{Some possible approaches to future research on genetic circuit design.}
    \label{tab:gaps}
\end{table}

 From the CCM, it can be seen that oscillators, repressilators and clocks are the most studied genetic circuits. The most explored design characteristic is retroactivity. This is understandable because retroactivity must be addressed for any genetic circuit to function as per design. However, an important observation is that handling retroactivity has been studied only in the context of oscillatory circuits, while its effect on other genetic circuits remains an open research question. Moreover, context-dependent effects such as noise and resource competition have not been as intensively studied as retroactivity, and these topics remain open for future work. Interestingly, two of the listed strategies for robust design have been applied for oscillatory circuits but not circuits with most other functionalities. A significant body of work has been done using ML approaches to design gene regulatory circuits. Still, a closer look reveals that the application of ML has to an extent been limited to the use of advanced optimisation algorithms rather than core ML methods. Therefore, there is much room for using ML in this area with some recent works exploring the same~\cite{Zhu2021}. Most of the design methodologies listed have been used to design oscillatory circuits, but they have not been utilised to design other types of circuits. Besides oscillatory circuits, only logic gates have received some attention from researchers.

The parameters and their corresponding options that form the CCM rows and columns were identified based on literature review. No review is exhaustive, and there is perhaps scope for adding new parameters and more options beyond the ones discussed herein. The current work intends to explore the research gaps using the CCM in Table~\ref{tab:ccm}. The blank cells in the CCM are potential areas for future research, but some may not be feasible. So the feasibility of the topics needs to be tested. Many approaches to further research in genetic circuit design can be identified easily with the help of the CCM. Some of the columns or rows, for instance, the pulse width modulator, have all blank cells since the existing work on it could not be matched to a suitable cell in the current CCM. 

The number of parameters listed in Table~\ref{tab:ccm} could be extended by finding new ones that underlie genetic circuit design. Furthermore, the options listed under each parameter may be extended and using this approach, some of the new research directions identified are: (a) finding novel design methodologies, (b) designing circuits with novel functionalities, (c) identifying potential design characteristics, and (d) finding new strategies for robust design. On the other hand, new directions can also be identified by probing the already identified parameters and options. This involves analysing the intersection of the rows and columns, such as (a) control-theoretic design methods may incorporate retroactivity and noise as design characteristics, (b) the separation of time-scale may be employed to make the design of a toggle switch more robust, (c) a list of topologies of robust circuits may be investigated using rule-based design, and (d) a memory circuit that is non-retroactive may be designed. Using a similar template as Table~\ref{tab:ccm}, some approaches to further research on genetic circuit design have been listed in Table~\ref{tab:gaps}. Although this is not a comprehensive list, the aim is to lay the foundation for a systematic method of generating research ideas using the generalised morphological analysis and CCM frameworks.

As scientists across the world chart the course for research in synthetic circuit design, reviews such as these can aid in systematically mapping out the contours of existing research, highlighting the current state-of-the art, as well as research gaps and various exciting possibilities. 

\subsection*{Acknowledgements}

The authors thank P. Sengupta for help with preparing the illustrations. DC acknowledges the HTRA fellowship from the Ministry of Human Resource Development, Government of India.

\bibliographystyle{abbrv}
\bibliography{References}
\end{document}